\documentclass[12pt]{iopart}

\usepackage{cite}
\usepackage[pdftex]{graphicx}
\usepackage{float}
\DeclareGraphicsExtensions{.pdf,.jpeg,.jpg,.png}
\usepackage{caption}
\usepackage{subcaption}
\usepackage{siunitx}

\begin{document}

\title[Fabrication of Surface Ion Traps with Integrated Current Carrying Wires]{Fabrication of Surface Ion Traps with Integrated Current Carrying Wires enabling High Magnetic Field Gradients}

\author{Martin~Siegele-Brown$^{1\dagger}$, Seokjun~Hong$^{1\dagger}$, Foni~Raphaël~Lebrun-Gallagher$^{1,2}$, Samuel~James~Hile$^{1}$, Sebastian~Weidt$^{1,2}$, and Winfried~Karl~Hensinger$^{1,2}$}% 
\address{$^{1}$Sussex Centre for Quantum Technologies, University of Sussex,\\ ~Brighton, BN1 9RH, UK}
\address{$^{2}$Universal Quantum Ltd, Brighton, BN1 6SB, UK}
\address{$^{\dagger}$These authors contributed equally to this work.}
\ead{W.K.Hensinger@sussex.ac.uk}

\vspace{10pt}
\begin{indented}
\item[] Keywords: microfabrication, surface ion traps, current carrying wires, magnetic field gradients, quantum computing
\end{indented}

\begin{abstract}
A major challenge for quantum computers is the scalable simultaneous execution of quantum gates. One approach to address this in trapped ion quantum computers is the implementation of quantum gates based on static magnetic field gradients and global microwave fields. 
In this paper, we present the fabrication of surface ion traps with integrated copper current carrying wires embedded inside the substrate below the ion trap electrodes, capable of generating high magnetic field gradients. The copper layer's measured sheet resistance of \SI{1.12}{\milli\ohm}/sq at room temperature is sufficiently low  to incorporate complex designs, without excessive power dissipation at high currents causing a thermal runaway. At a temperature of \SI{40}{K} the sheet resistance drops to \SI{20.9}{\micro\ohm}/sq giving a lower limit for the residual resistance ratio of 100. Continuous currents of \SI{13}{A} can be applied, resulting in a simulated magnetic field gradient of \SI{144}{T/m} at the ion position, which is \SI{125}{\micro\meter} from the trap surface for the particular anti-parallel wire pair in our design.

\end{abstract}

\section{Introduction}

Since the first quantum gates using trapped ions were proposed by Cirac and Zoller \cite{cirac1995quantum}, trapped ions have been considered to be a promising qubit platform to realise large-scale quantum information processing. Experiments have proven the capabilities of trapped ions to be used as physical qubits, including high-fidelity state preparation \cite{Wunderlich2007,Blatt2008,Harty2014}, universal gate operation \cite{Monroe1995,Harty2014,Akerman2015} and readout \cite{Noek2013,Harty2014}, and long coherence times \cite{timoney2011, Harty2014, wang2021single}. In addition to these experiments using single ion strings, there has also been research into expanding the ion trapping system into a large scale architecture, and increasing the number of qubits that can be processed in parallel and entangled for implementation of more complex quantum algorithms \cite{brown2016co, bruzewicz2019trapped}. The key of this research, inspired by the proposals of Wineland et al.~\cite{wineland1998experimental} and Kielpinski et al.~\cite{kielpinski2002}, is making all macroscopic devices for trapping ions completely scalable. Surface ion traps developed using micro-electro-mechanical system (MEMS) fabrication technology \cite{stick2006ion, Seidelin, Hughes2011 ,romaszko2020engineering} and the concept of junction nodes connecting linear ion trapping zones \cite{Hensinger, amini2010toward, wright2013reliable} have significantly increased the number of ions that can be trapped and controlled on a microchip.  Optical and electrical components such as micro-mirrors \cite{merrill2011demonstration} and optical waveguides \cite{mehta2016integrated, mehta2020integrated, niffenegger2020integrated, ivory2021integrated}, photodetectors \cite{Eltony,Todaro2021, setzer2021fluorescence}, digital-to-analogue converters (DACs) \cite{stuart2019chip}, and trench capacitors \cite{guise2015ball} can be fabricated in microchips and integrated in the same chip together with ion traps.

A major challenge for increasing the number of qubits is the simultaneous large-scale control of quantum gates, and a few methods have been developed for trapped ions. For laser based gates, integrated optical waveguides and gratings \cite{mehta2016integrated, mehta2020integrated, niffenegger2020integrated, ivory2021integrated} are a promising approach. 
As a different approach, laser free high fidelity gates have been demonstrated using oscillating magnetic field gradients \cite{zarantonello2019robust,srinivas2021high}.
Another approach is to use long-wavelength radiation combined with a static magnetic field gradient as proposed by Mintert and Wunderlich \cite{Mintert2001}.
Weidt et al.~have further built on this work and demonstrated a quantum computing approach where quantum gates can be executed simply by the application of a semi-static voltage to a microchip making use of locally applied magnetic field gradients, and global microwave and RF fields \cite{Weidt}. Lekitsch et al.~have developed a microchip-based blueprint for a fault-tolerant quantum computer capable of hosting millions of qubits \cite{Lekitsche1601540}. In the proposed architecture, microfabricated surface ion traps and integrated current-carrying wires (CCWs) are used to trap ions and generate a high static magnetic field gradient in designated gate zones on the chip.

Historically, CCWs have been used extensively for the confinement of neutral atoms, allowing the technology to shift from free-standing wires to more robust microfabricated atom chips \cite{FOLMAN2002263}. 
In atom chips, a single layer of evaporated gold up to several micrometres thick is widely used for CCWs \cite{folman2000controlling, groth2004atom}.
Furthermore, multilayer atom chips have been demonstrated \cite{trinker2008multilayer}, some allowing currents up to \SI{10}{A} in \SI{15}{\micro\metre} thick CCWs \cite{chuang2011design}, including one featuring \SI{2}{\micro\metre} thick copper CCWs embedded into the silicon \cite{chuang2011multi}.
For ion traps with integrated current carrying wires, the wires need to be combined with a top layer structure capable of applying the required high RF voltages, and dielectrics in sight of the ion should be avoided.

In ion traps with CCWs in the top layer, static axial magnetic field gradients of 2.3--\SI{23}{T/m} have been demonstrated \cite{Wang2009, welzel2011designing, kunert2014planar}. 
Oscillating magnetic field gradients of 7, 35 and \SI{55}{T/m} in the 1-\SI{3.2}{GHz} range have been reported at ion heights of 75, 30 and \SI{35}{\micro\metre}  respectively \cite{harty2016high, warring2013techniques, hahn2019multilayer}, as well as a radial gradient at 5 MHz of 152(15) T/m at an ion height of \SI{30}{\micro\metre} \cite{srinivas2021high}, all with the CCWs located in the top layer of the ion trap.
In a different implementation where \SI{127}{\micro\meter} thick copper wires have been placed on a chip carrier below the ion trap, a static magnetic field gradient of 16 T/m at an ion height of \SI{96.9}{\micro\metre} has been achieved \cite{Welzel_2018}. %96.9um ion height

In our design the CCWs are embedded in the silicon substrate perpendicular to the RF trapping electrodes as shown in Figure 1. In this concept, the design of the CCWs is completely independent from the design of the top electrode layer. This removes the constraints that have limited the axial gradient with CCWs in the top layer, without moving the CCWs significantly further away from the ion. With a sufficiently low sheet resistance of the copper for the management of thermal dissipation from the current, high currents in excess of \SI{10}{A} can be applied allowing magnetic field gradients in the range of 100--\SI{150}{T/m} at an ion position \SI{125}{\micro\metre} from the trap surface.

\begin{figure}[t!]
\centering
\includegraphics[width=\textwidth]{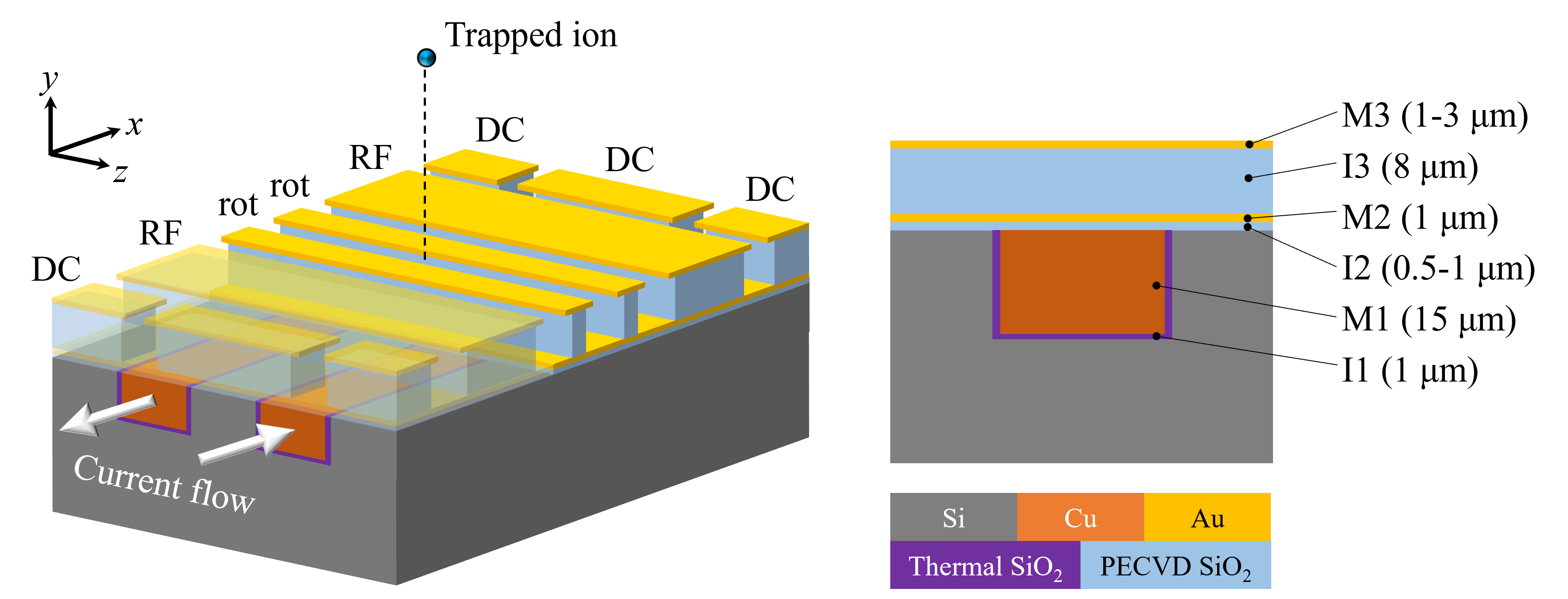}
\caption{Schematic of the proposed surface ion trap with integrated CCWs (not to scale). A set of segmented DC electrodes and RF electrodes are shown as transparent to show part of the copper wires integrated under the electrodes. The vertical dimensions of the ion trap electrodes and CCWs are shown on the right.}
\label{fig:Schematic}
\end{figure}

In this paper, we present the design and fabrication of surface ion traps with integrated CCWs for high magnetic field gradients. A fabrication process consisting of integration of copper wires in the silicon substrate and fabrication of ion trap electrodes on top of this structure was developed. In order to reduce the effect of stray electric fields induced by built-up charges on dielectric surfaces, gold was used for conducting layers exposed to the ion to avoid native oxide formation, and the thick oxide layer underlying the uppermost metal layer was undercut. Application of high currents up to \SI{13}{A} continuous, resulting in a simulated magnetic field gradient of \SI{144}{T/m} at the ion position of \SI{125}{\micro\meter}, was demonstrated on the fabricated wires, and power dissipation was measured. The fabricated chips were also used to trap $^{171}$Yb$^+$ ions. 
Considering the gate mechanism, a magnetic field gradient of \SI{120}{T/m} should enable much stronger sideband transitions and is predicted to give rise to a two-qubit gate fidelity of $\sim$\SI{99.9}{\percent} \cite{Weidt}, significantly higher than the fault-tolerant threshold for the surface code \cite{fowler2012surface}.

\section{Design}

The vertical dimensions of the designed ion trap electrodes are shown in Figure 1. The thickness of the oxide layer (I3) isolating the top layer (M3) and the ground plane (M2) is \SI{8}{\micro\meter}, where we measured a flashover voltage of over \SI{250}{\volt_{amp}} at \SI{15}{MHz} between the RF electrode and the ground plane using a test structure chip. 
The ion height numerically simulated by the lateral design of electrodes is \SI{125}{\micro\meter} from the electrode surface. The two central DC electrodes between the RF rails are used to apply compensating DC voltages for principal axis rotation. 

\begin{figure}[t!]

\centering
\includegraphics[width=0.8\textwidth]{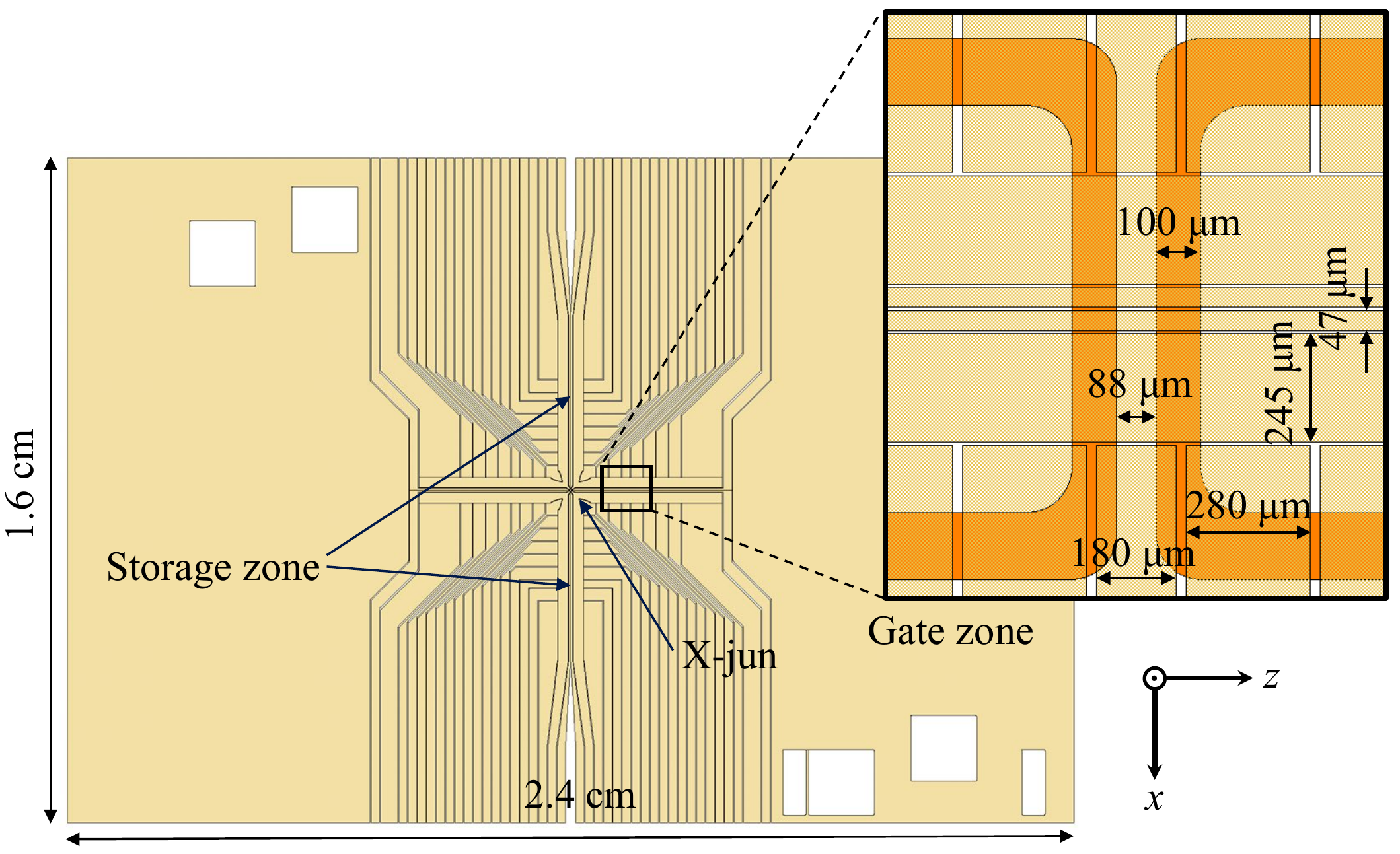}
        \caption{Layout design of ion trap showing the gate zone and storage zones. Ions can be shuttled between zones via the X-junction in the centre of the device. The white squares represent the bonding pads for the CCWs. The magnified view shows the electrode and CCW design in the gate zone.} 
        \label{fig:}
\end{figure}

\begin{figure}[t!] %change to !t
     \centering
     \begin{subfigure}[b]{0.49\textwidth}
\centering
\includegraphics[width=\textwidth]{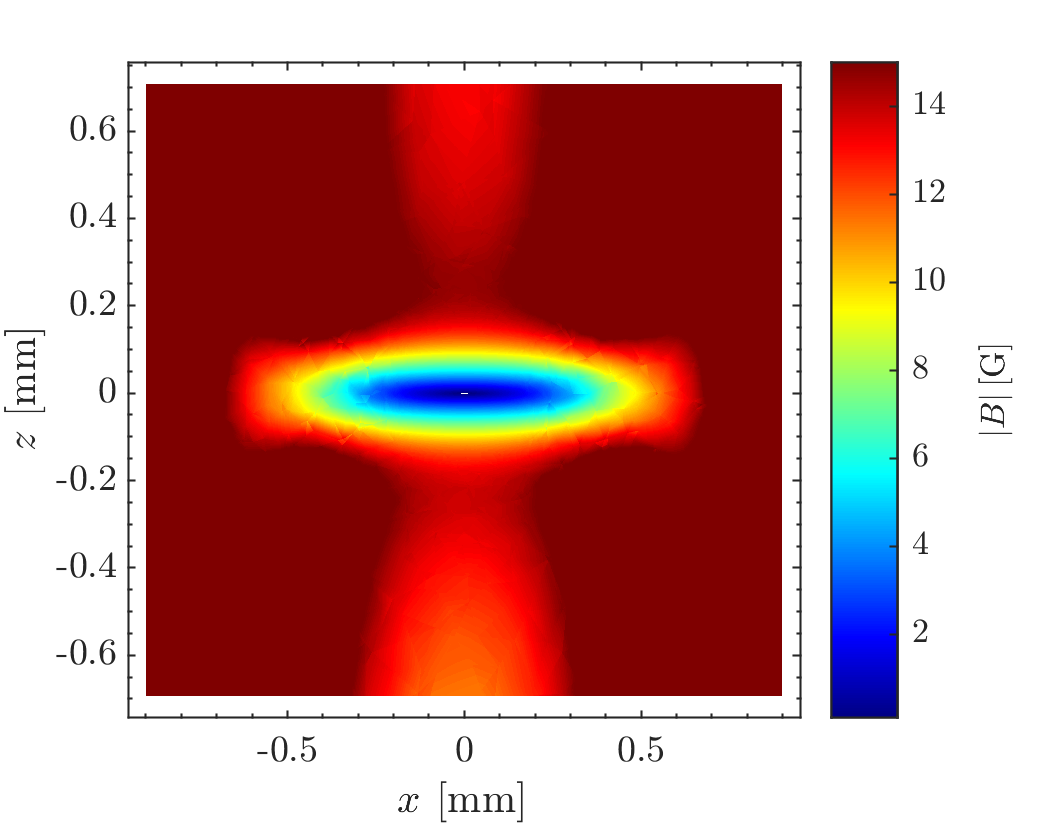}
\caption{}
     \end{subfigure}
     \hfill
     \begin{subfigure}[b]{0.49\textwidth}
\centering
\includegraphics[width=\textwidth]{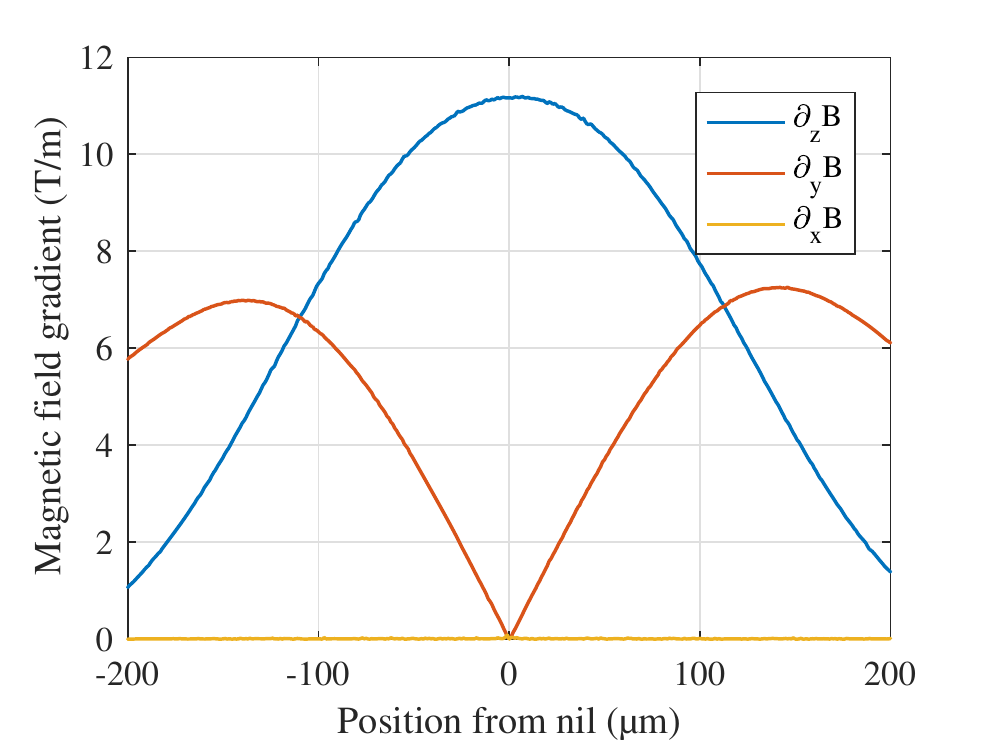}
\caption{}
\end{subfigure}
\caption{Finite element method simulation results of the CCW structure designed for this experiment per \SI{1}{A} input current. (a) Magnetic field magnitude $|B|$ in a horizontal plane at the trapping height ($h_{ion}=\SI{125}{\micro\meter}$) with the magnetic field quadrupole clearly visible in the centre at (x = 0; y = $h_{ion}$; z = 0). (b) x-, y-, and z-components of the gradient of the magnetic field magnitude along the axial trapping axis z. Close to the quadrupole centre, the magnetic field gradient $\partial B$ is maximal with a gradient of \SI{11.1}{T/m} per \SI{1}{A} input current. At \SI{13.4}{A} this gives the \SI{150}{T/m} necessary for high fidelity gates proposed by Weidt et al.~\cite{Weidt}.}

\label{fig_}
\end{figure}

The gate zone incorporating the CCWs is located in one of the four arms of the X-junction (Figure 2). The design of the X-junction is adapted from \cite{mokhberi2017optimised}. The trapped ions can be shuttled to two storage zones with a low magnetic field of $<\SI{2}{mT}$ placed in the two perpendicular arms. Two parallel wires are placed perpendicular to the RF rails, and DC currents with opposite directions are applied to the wires to maximise the magnetic field gradient at the ion position. 
\SI{100}{\micro\meter}, \SI{15}{\micro\meter}, and \SI{88}{\micro\meter} for the width and depth of wires and the distance between the wires respectively were chosen to give a good trade-off between current density and power dissipation, and maximum magnetic field gradient. The gradient is only reduced by \SI{17}{\percent} compared to the ideal case of two infinitely thin wires located on the silicon surface, while limiting the current density to \SI{1e6}{A\per\centi\square\meter} at \SI{15}{A}.
The overall CCW structure corresponds to a wire with a length to width ratio of 390. Finite element method simulation results show that by applying \SI{13.4}{A} to the wires, a magnetic field gradient of \SI{150}{T/m} along the longitudinal trap axis can be generated at the ion position (Figure 3). Depending on the precise gate implementation, a gradient in the range of 100--\SI{150}{T/m} is desirable.

A further consideration for the CCW design is ohmic power loss and substrate heating, mainly determined by the resistance of the wires. The \SI{15}{\micro\meter} thick copper layer has a sheet resistance of \SI{1.12}{\milli\ohm}/sq at room temperature. 
This low sheet resistance allows for a resistance at room temperature of \SI{438}{\milli\ohm} for our CCWs in series, despite the complex design.
The resistance is predicted to be reduced to 15.0--\SI{49.5}{\milli\ohm} at the intended operating temperature of 40--\SI{70}{K}, assuming a copper layer with a residual resistance ratio (RRR) of 50. 
This resistance corresponds to a power dissipation of \SI{2.3}{W} when applying \SI{10}{A} assuming a cooling system with a base temperature of 40~K and 5~K/W overall thermal resistance, which is a conservative estimate for our setup from previous measurements.

\section{Fabrication}

\begin{figure}[t!]
     \centering
     \begin{subfigure}[b]{0.3\textwidth}
         \centering
         \includegraphics[width=\textwidth]{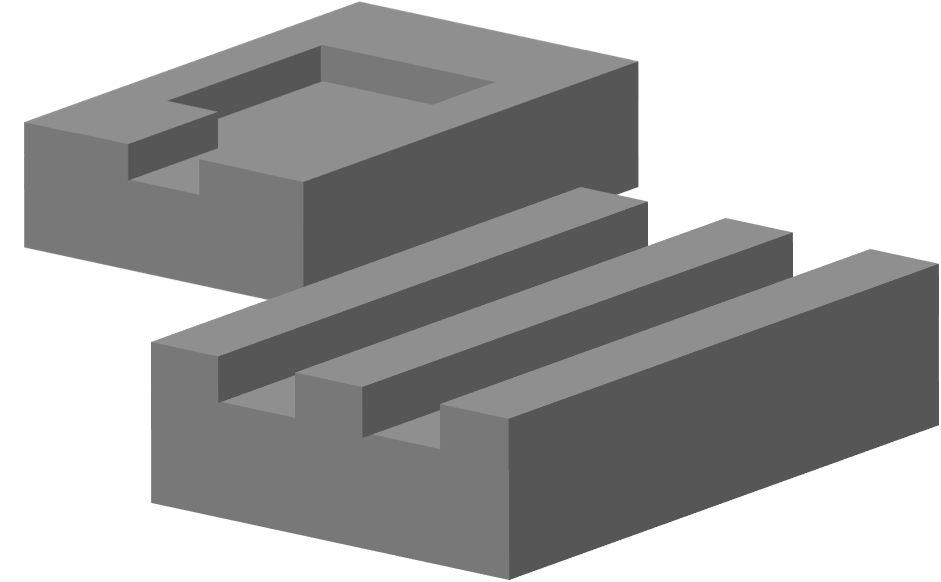}
         \caption{Si dry etch}
         \label{fig:pf_a}
     \end{subfigure}
     \hfill
     \begin{subfigure}[b]{0.3\textwidth}
         \centering
         \includegraphics[width=\textwidth]{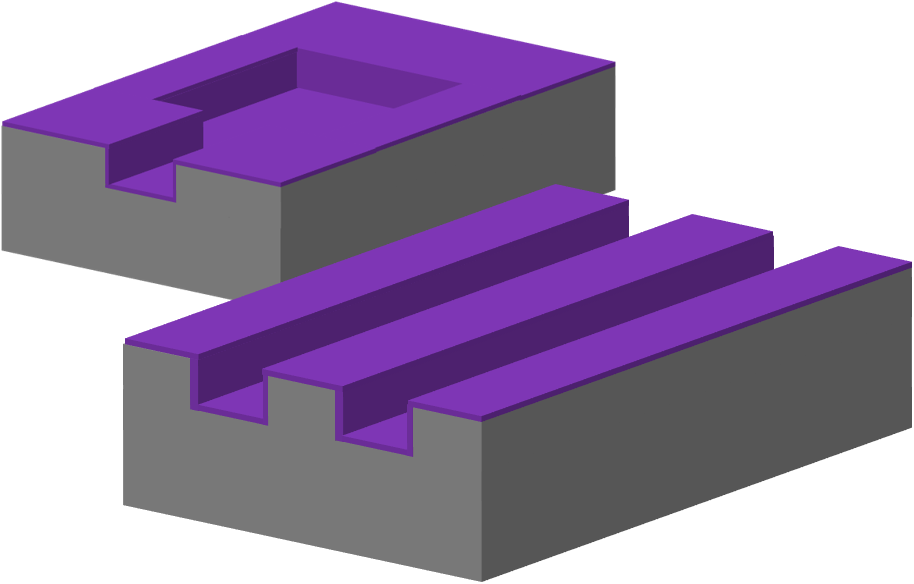}
         \caption{Thermal oxidation}
         \label{fig:pf_b}
         \end{subfigure}
    \hfill
     \begin{subfigure}[b]{0.3\textwidth}
         \centering
         \includegraphics[width=\textwidth]{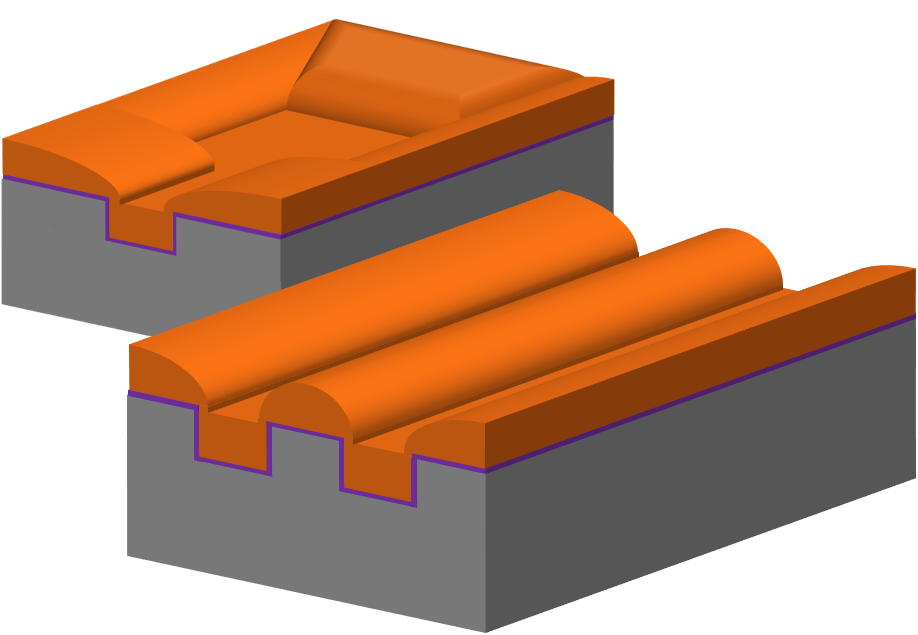}
         \caption{Ti/Cu dep. \& Cu electroplating}
         \label{fig:pf_c}
     \end{subfigure}
         \hfill
     \begin{subfigure}[b]{0.3\textwidth}
         \centering
         \includegraphics[width=\textwidth]{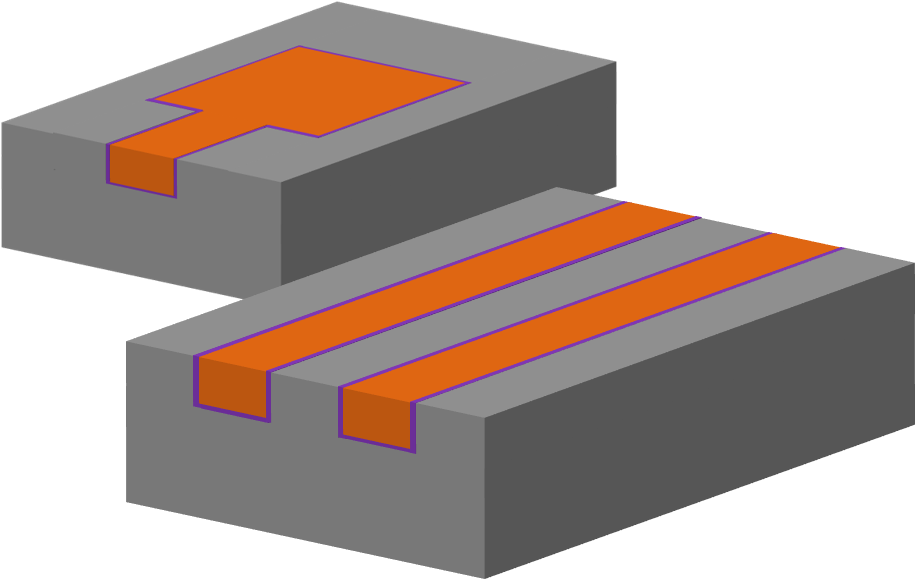}
         \caption{CMP}
         \label{fig:pf_d}
     \end{subfigure}
              \hfill
     \begin{subfigure}[b]{0.3\textwidth}
         \centering
         \includegraphics[width=\textwidth]{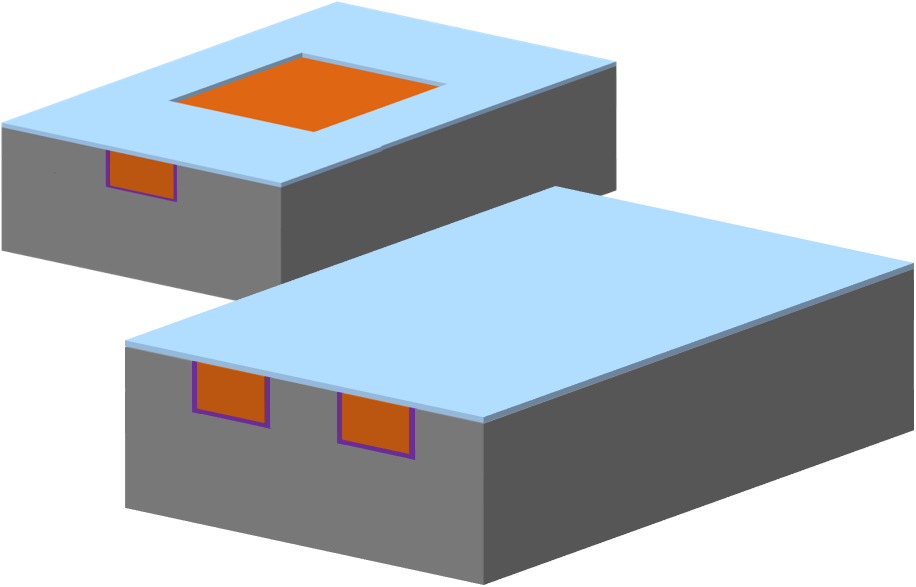}
         \caption{SiO$_2$ dep. \& wet etch}
         \label{x}
     \end{subfigure}
              \hfill
     \begin{subfigure}[b]{0.3\textwidth}
         \centering
         \includegraphics[width=\textwidth]{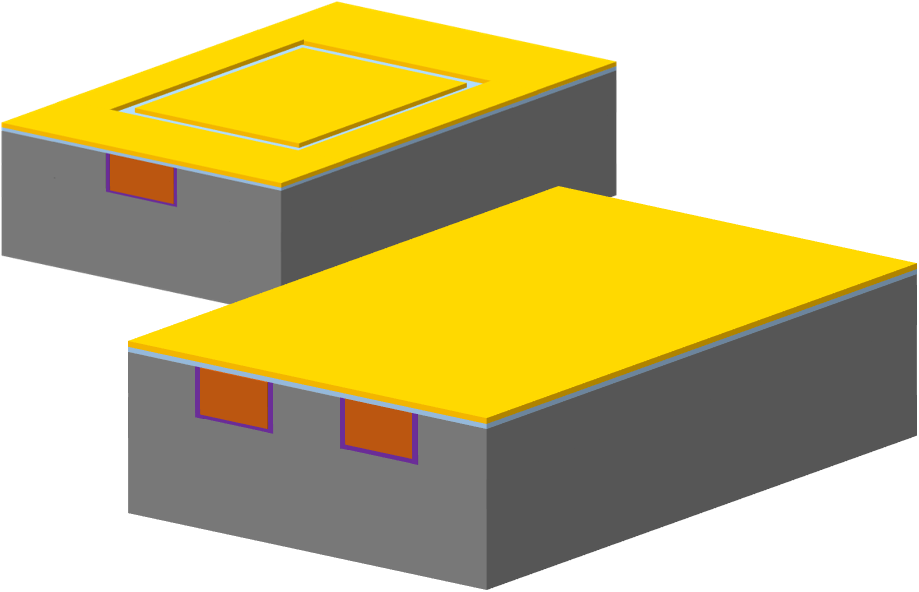}
         \caption{Cr/Au/Cr dep. \& dry etch}
         \label{x}
     \end{subfigure}
              \hfill
     \begin{subfigure}[b]{0.3\textwidth}
         \centering
         \includegraphics[width=\textwidth]{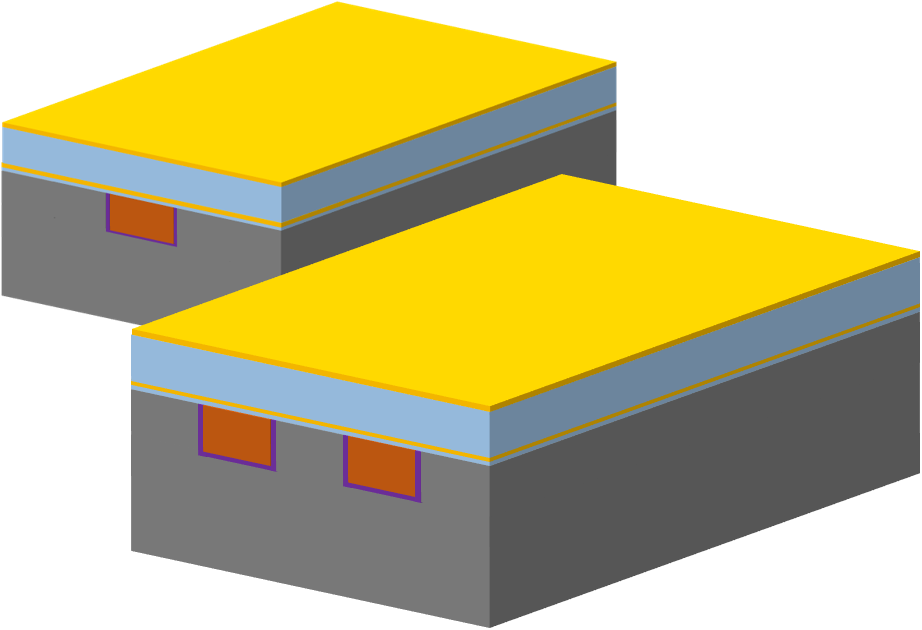}
         \caption{SiO$_2$/Cr/Au dep.}
         \label{x}
     \end{subfigure}
              \hfill
     \begin{subfigure}[b]{0.3\textwidth}
         \centering
         \includegraphics[width=\textwidth]{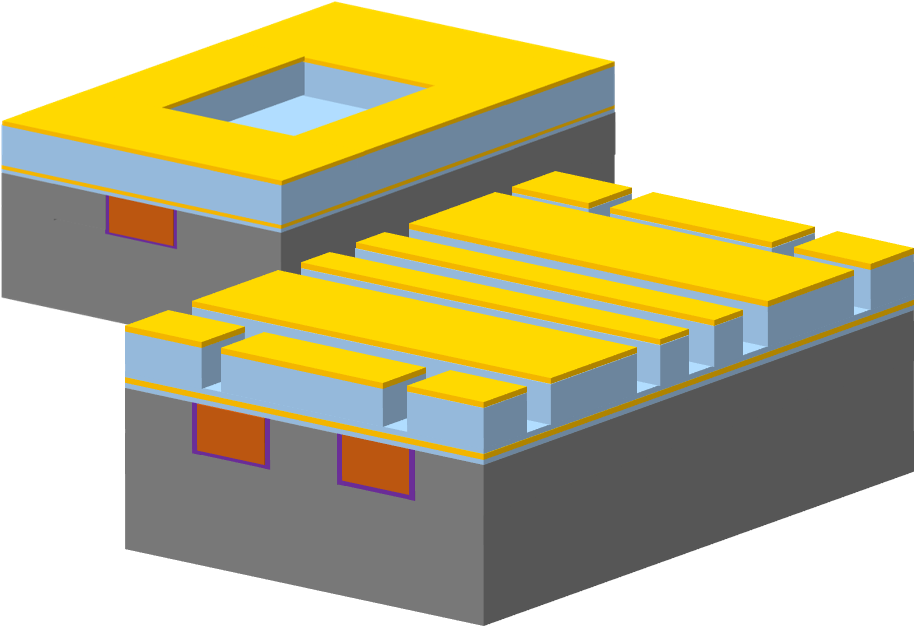}
         \caption{Au/Cr/SiO$_2$ dry etch}
         \label{x}
     \end{subfigure}
              \hfill
     \begin{subfigure}[b]{0.3\textwidth}
         \centering
         \includegraphics[width=\textwidth]{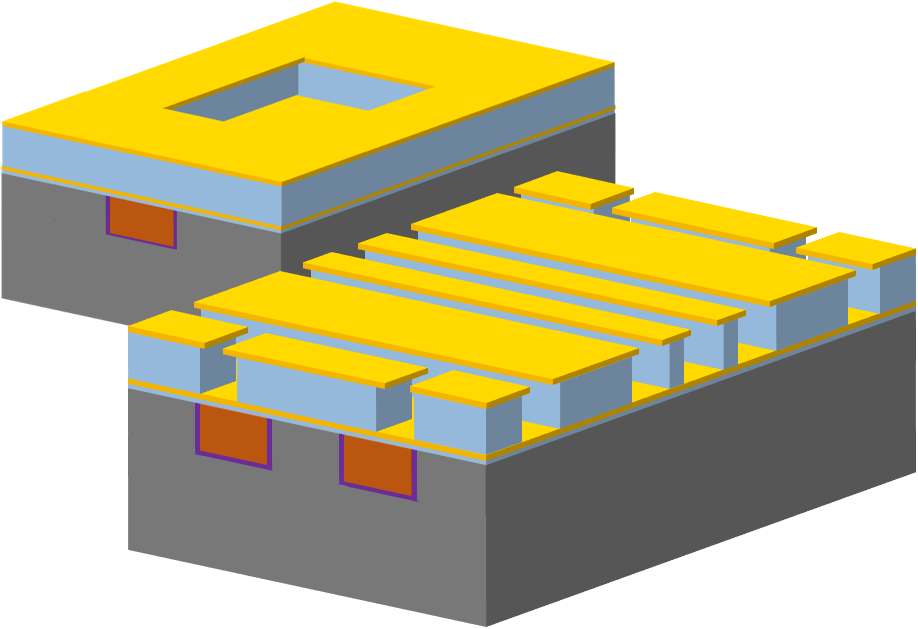}
         \caption{SiO$_2$/Cr wet etch}
         \label{x}
     \end{subfigure}
        \caption{Fabrication process flow. Note that the adhesion layers and photolithography steps are not shown in the schematics. Positive photoresist was used for all the lithography steps. The schematics at the front and back show the process flow of the gate zone area and bonding pads of the CCWs respectively. All schematics not to scale.}
        \label{fig:}
\end{figure}

This section describes the fabrication process of integrating the CCWs into the silicon substrate and building ion trap electrodes on top of the CCW-integrated wafer. 
First, the CCW layout is photolithographically patterned on the silicon wafer using a conventional deep reactive ion etching (DRIE) process (Figure 4(a)). After stripping the used resist followed by a standard RCA clean, the wafers are thermally oxidised in a furnace tube to grow wet oxide of \SI{1}{\micro\meter}, forming an insulating layer between the silicon substrate and the CCW (Figure 4(b)). Then, Ti/Cu with a thickness of 10/\SI{200}{nm} is deposited by a sputtering process, and using this as an adhesion and seed layer, a \SI{25}{\micro\meter}-thick copper layer is subsequently electroplated to form the CCWs (Figure 4(c)). The fabricated structures are planarised by a three-step sequence that consists of diamond milling and two chemical mechanical polishing (CMP) steps. 
Optimising the process parameters, a dishing of approximately \SI{100}{\nano\meter} for \SI{100}{\micro\meter} to \SI{1500}{\micro\meter} wide patterns was achieved.
This concludes the fabrication of the CCW base wafer (Figure 4(d)).

\begin{figure}[b!]
\centering
\includegraphics[width=\textwidth]{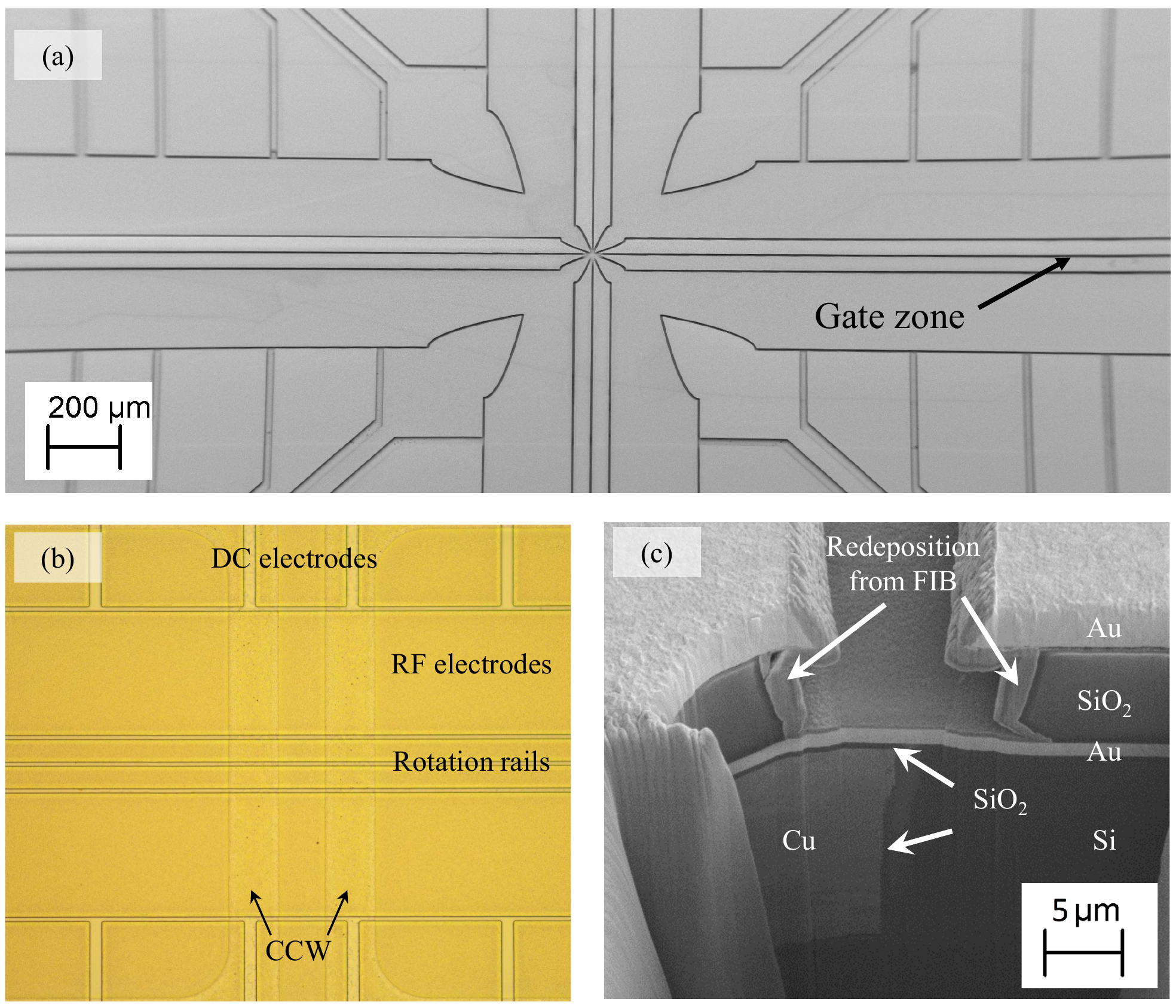}
\caption{(a) SEM image of the X-junction. (b) Optical microscope image of the trap, showing the gate zone. (c) SEM image of the cross-section of the device made by a focused ion beam.}
\label{fig:sems}
\end{figure}

Fabrication of the ion trap electrodes starts with plasma-enhanced chemical vapour deposition (PECVD) of a 0.5--\SI{1}{\micro\meter}-thick SiO$_2$ layer (Figure 4(e)). The oxide layer is patterned by wet etching using buffered oxide etchant (BOE) (HF:NH$_4$F = 1:7) to open the bonding pads for the CCWs. Then, a Cr/Au/Cr layer of 10/1000/\SI{10}{nm} is sputtered and ion-beam-etched (IBE) to provide the ground plane as well as the bonding pads for the CCWs (Figure 4(f)). The thin chromium layer is used to increase adhesion between silicon oxide and gold. Next, an \SI{8}{\micro\meter}-thick SiO$_2$ layer and 10/1000--\SI{3000}{\nano\meter}-thick Cr/Au layer are deposited (Figure 4(g)). These two layers are dry etched sequentially using the same photoresist mask (Figure 4(h)). The good uniformity (4(1)\%) of the deep oxide etching process allows stopping \SI{750(250)}{nm} before reaching the Cr/Au/Cr layer, in order not to expose the gold to the oxide etching chemistry. The remaining oxide is etched by BOE based wet etching. This wet etching also etches the sidewalls of the oxide pillar, resulting in the formation of the electrode undercut structure (Figure 4(i)). After the BOE, the exposed Cr adhesive layer over the Au ground plane is removed by an additional wet etching process. Finally, a 10/\SI{500}{nm} Cr/Au layer is sputtered on the back of the wafer to improve indium bonding of the traps.

An optical microscope image showing the top view of the gate zone, a scanning electron image (SEM) image of the X-junction, and an SEM image of the cross-section of the devices fabricated by a focused ion beam (FIB) are shown in Figure \ref{fig:sems}. The cross-section is placed at at the end of the west arm of the X-junction where a routing wire of the CCW is parallel to the two rotation electrodes, which are separated by \SI{5}{\micro\meter}. The oxide below the rotation electrodes shows the desired undercut of \SI{3(1)}{\micro\meter} from the oxide wet etch.
Every surface in the line of sight of the ion beam impact is covered with some redeposited material, which is unavoidable.
The resistance between the copper tracks and surrounding conducting layers was higher than our measurement limit (\SI{10}{\mega\ohm}), and no unintended electrical short was observed between ion trap electrodes.

\section{Experiments}

\begin{figure}[b!]
\centering
\includegraphics[width=\textwidth]{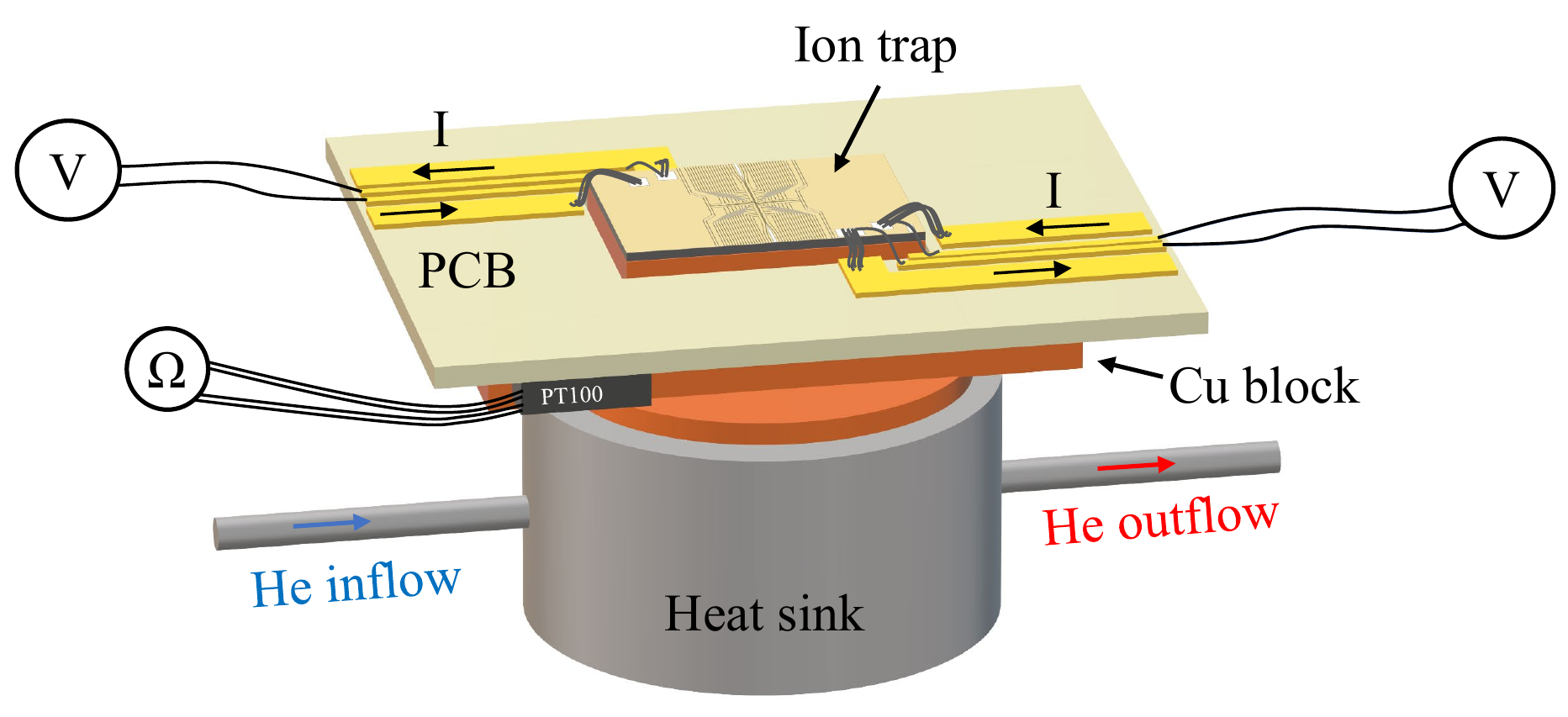}
\caption{The fabricated trap is bonded to a copper block using indium with a PCB attached. The CCWs are connected by ten 25$\times$\SI{125}{\micro\meter} aluminium ribbon bonds per pad to a printed circuit board (PCB) and by one ribbon bond for voltage sensing connections. The temperature is measured using a PT100 attached to the copper block. The copper block is then mounted on a heat sink located in a custom UHV test chamber, which is connected to a custom closed-loop cryogenic helium gas circulation system.
}
\label{fig:Schematic}
\end{figure}

The resistance of the CCW at room temperature was measured using four-terminal sensing with \SI{100}{mA} source current. The fabricated CCWs match the expected resistance within $<$1\%.
To evaluate performance at low temperatures, the device was mounted in a custom cryogenic system as shown in Figure \ref{fig:Schematic}. 
The fabricated trap was indium soldered to a copper block, which extends through a hole in the printed circuit board (PCB) to which it is attached. The CCWs were connected by ten 25$\times$\SI{125}{\micro\meter} ribbon bonds per pad to the PCB and by one ribbon bond for voltage sensing connections. The temperature was measured with a calibrated PT100 (Lakeshore Pt-111-14H) attached to the copper block. The copper block was then mounted on a heat sink (using AuSn preforms in-between) located in a custom ultra high vacuum (UHV) test chamber, which was supplied by a custom closed loop cryostat \cite{lebrun2021scalable}. The measurements were performed at a pressure of $\sim10^{-6}$~mbar and a base temperature of \SI{38(2)}{K}.

\begin{figure}[b!]
     \centering
     \begin{subfigure}[b]{0.49\textwidth}
         \centering
         \includegraphics[width=\textwidth]{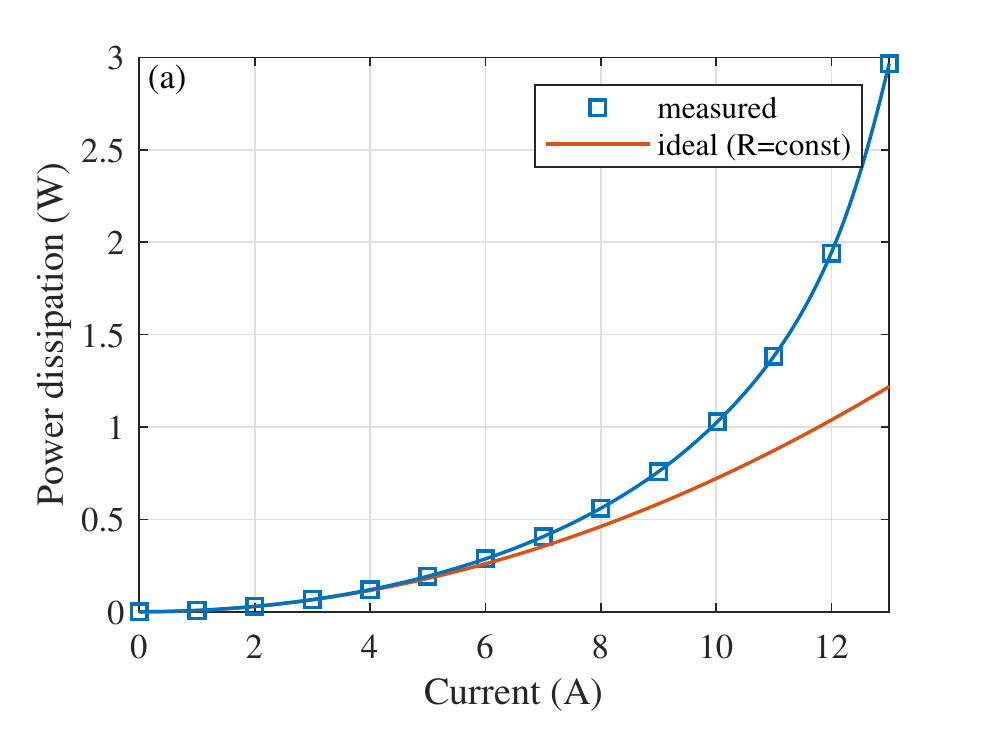}
         \label{fig:meas_P}
     \end{subfigure}
     \begin{subfigure}[b]{0.5\textwidth}
         \centering
         \includegraphics[width=\textwidth]{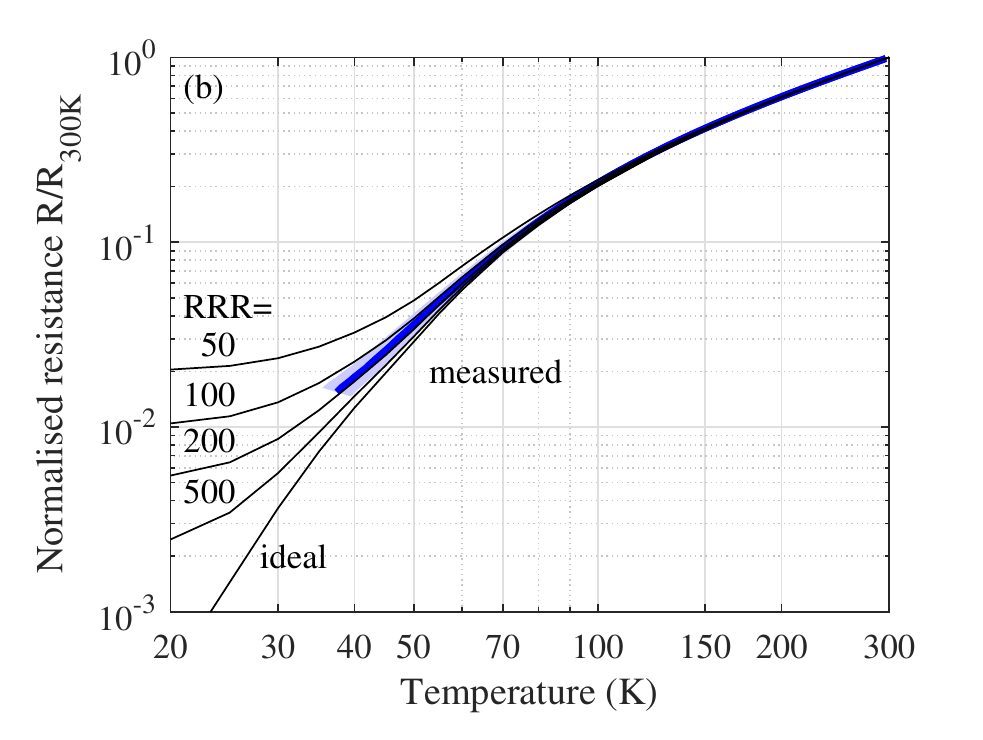}
         \label{fig:meas_R}
     \end{subfigure}
        \caption{(a) Measured power dissipation over applied current through the whole CCW structure, compared to the ideal case where the trap does not heat up. (b) Normalised resistivity of CCW over temperature. The measured resistance (blue with the uncertainty as shaded area) is compared to copper with different RRR for comparison.}
        \label{fig:}
\end{figure}

To characterise the power consumption of the CCWs, current was applied through all CCWs in series, and the voltage drop was directly measured at the CCW pads (via sensing bond wires). At a current of \SI{10}{A} the power dissipation is \SI{1028(10)}{mW} (Figure 7(a)), which is 1.43 times higher than it would be without heating the trap. The rise in resistivity by a factor of 1.43 corresponds to a rise in temperature of the CCWs from \SI{38}{K} to \SI{43}{K}, while the temperature measured on the copper block is \SI{40}{K}. The maximum continuous current of \SI{13}{A} was limited by thermal runaway which can be improved by optimising the thermal anchoring. 

To obtain the temperature dependence of the resistance, the helium flow was turned off and the resistance was measured as the system was heating up (\SI{2}{K/min}). A probe current of \SI{10}{mA} was used for the four-terminal sensing of the CCW resistance. At \SI{40(1)}{K} the resistance drops from \SI{438(1)}{\milli\ohm} by a factor of 54(2) to \SI{8.2(4)}{\milli\ohm} and at \SI{70(1)}{K} by a factor of 10.7(1) to \SI{40.8(4)}{\milli\ohm}. This corresponds to a sheet resistance of \SI{20.9(9)}{\micro\ohm}/sq at \SI{40(1)}{K} and \SI{104(1)}{\micro\ohm}/sq at \SI{70(1)}{K}.  Figure 7(b) shows the normalised resistance (blue with uncertainty as shaded area) compared to Cu with an RRR of 50, 100, 200 and 500 and pure Cu (data from \cite{matula1979electrical}). The measurement setup does not allow a precise number for RRR to be obtained, but is estimated as $180^{+215}_{-65}$ with a lower bound of 100.

\begin{figure}[t!]
\centering
\includegraphics[width=3in]{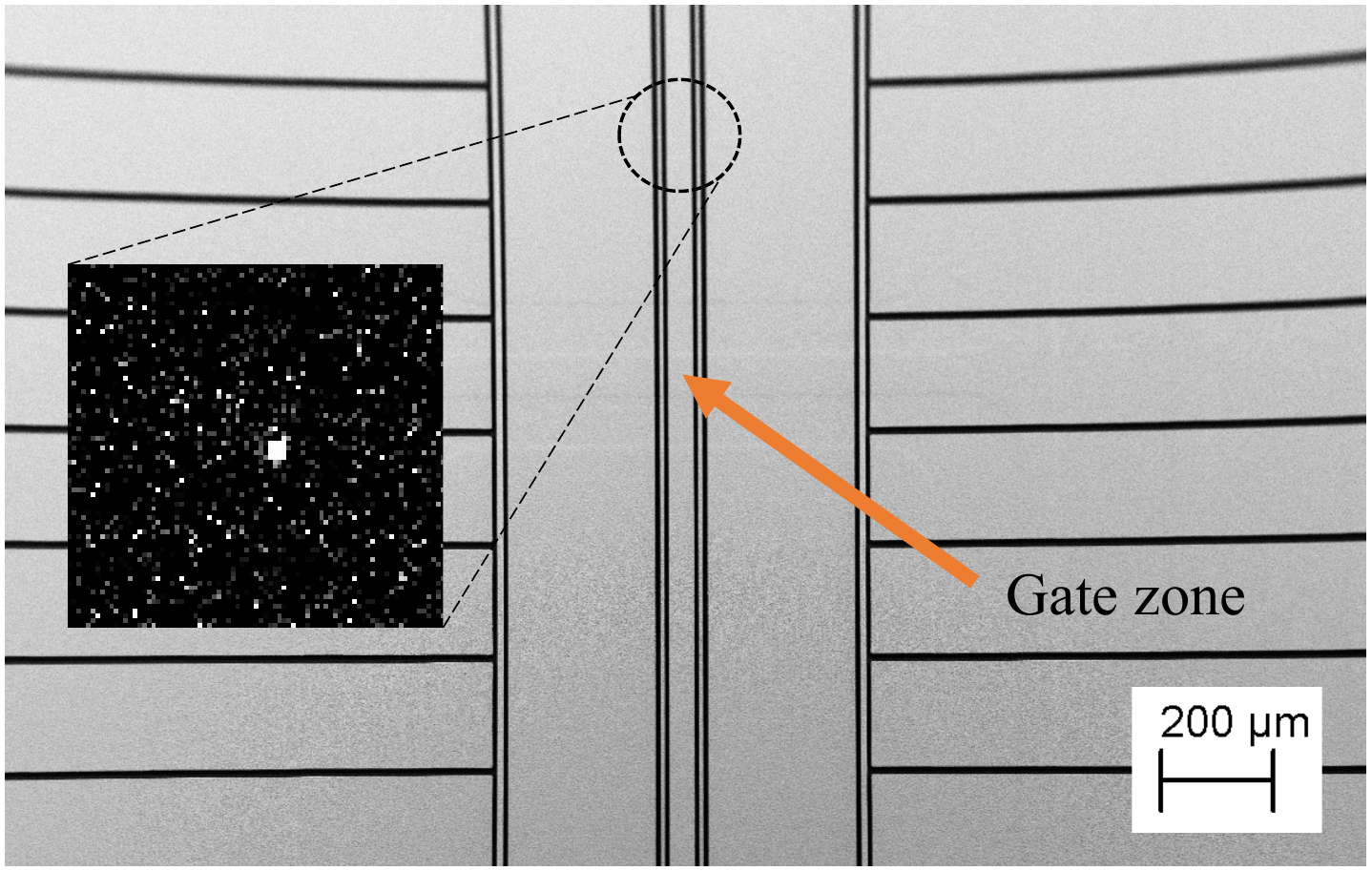}
\caption{SEM image of the fabricated device. The inlay shows an image of a trapped ion captured by an electron-multiplying charge-coupled device (EMCCD) sensor.}
\label{fig:ion}
\end{figure}

Fabricated chips with integrated CCWs were successfully used to trap $^{171}$Yb$^+$ ions as shown in Figure \ref{fig:ion}. 
We note that the chip used for trapping is slightly modified earlier version of the design described in Sec. 3 featuring an aluminium ground plane instead of a gold ground plane. We had adjusted the fabrication process in order to avoid an aluminium electrode, as such an electrode could potentially form a very thin oxide layer and charge up. We also reduced the number of central DC electrodes located in between the two RF rails for the newer design to enable a simpler principal axis rotation and to enable more undercut of the dielectric below the electrodes. During trapping, we note that we did not observe any adverse effects because of the aluminium ground plane.
Ions have been trapped on this device, frequently for 7--\SI{16}{h}. 

\section{Conclusion and outlook}

We have presented the fabrication of surface ion traps with current carrying wires integrated into the silicon substrate. The dishing of the CCWs is well controlled, with \SI{100}{nm} for both small features and large features, ensuring the electrical performance is not compromised. The residual resistance ratio is estimated as $180^{+215}_{-65}$ with a lower limit of 100 giving a resistance close to that of pure copper at the temperature ranges in our application.
These devices are capable of providing high currents of \SI{13}{A} continuous for high static magnetic field gradients of \SI{144}{T/m}, at an ion position of \SI{125}{\micro\meter} from the trap surface. These results indicate that the devices are suitable for high fidelity quantum gates based on static magnetic field gradients and global microwave fields, constituting a promising approach for building practical trapped ion quantum computers with millions of qubits. 
\section*{Acknowledgment}

The authors would like to thank Knut Gottfried, Cyrille Hibert, Didier Bouvet and Joffrey Pernollet for helpful discussions about the fabrication process, Reuben Puddy and Zak Romaszko for work on the early designs, and Mariam Akhtar and Falk Bonus for performing the trapping runs. Work was carried out at a number of facilities including the Center of MicroNanoTechnology (CMi) at École polytechnique fédérale de Lausanne (EPFL), the London Centre for Nanotechnology (LCN) and the Scottish Microelectronics Centre (SMC) at the University of Edinburgh.

This work was supported by the U.K. Engineering and Physical Sciences Research Council via the EPSRC Hub in Quantum Computing and Simulation (EP/T001062/1), the U.K. Quantum Technology hub for Networked Quantum Information Technologies (No. EP/M013243/1), the European Commission’s Horizon-2020 Flagship on Quantum Technologies Project No. 820314 (MicroQC), the U.S. Army Research Office under Contract No. W911NF-14-2-0106 and Contract No. W911NF-21-1-0240, the Office of Naval Research under Agreement No. N62909-19-1-2116, the Luxembourg National Research Fund (FNR) (Project Code 11615035), and the University of Sussex.

\section*{References}

\bibliographystyle{iopart-num}
\bibliography{references}

\end{document}